# Strong gravitational lensing probes of the particle nature of dark matter

## Submitted to the Astro2010 Decadal Cosmology and Fundamental Physics Science Frontier Panel

**Leonidas A Moustakas (JPL/Caltech), Kevork Abazajian (Maryland), Andrew Benson (Caltech), Adam S Bolton (IfA, Hawaii), James S Bullock (UC Irvine), Jacqueline Chen (Bonn), Edward Cheng (Conc. Analytics), Dan Coe (JPL/Caltech), Arthur B Congdon (JPL/Caltech), Neal Dalal (CITA, Toronto), Juerg Diemand (UCSC), Benjamin M Dobke (JPL/Caltech), Greg Dobler (CfA, Harvard), Olivier Dore (CITA, Toronto), Aaron Dutton (UCSC), Richard Ellis (Caltech), Chris D Fassnacht (UCD), Henry Ferguson (STScI), Douglas Finkbeiner (CfA, Harvard), Raphael Gavazzi (IAP), Fredrick William High (CfA, Harvard), Tesla Jeltema (UCSC), Eric Jullo (JPL/Caltech), Manoj Kaplinghat (UC Irvine), Charles R Keeton (Rutgers), Jean-Paul Kneib (Ld'A, Marseille), Leon V E Koopmans (Kapteyn), Savvas M Koushiappas (Brown), Michael Kuhlen (IAS, Princeton), Alexander Kusenko (UCLA), Charles R Lawrence (JPL/Caltech), Abraham Loeb (CfA, Harvard), Piero Madau (UCSC), Phil Marshall (UCSB), R Ben Metcalf (MPIA), Priya Natarajan (Yale), Joel R Primack (UCSC), Stefano Profumo (UCSC), Michael D Seiffert (JPL/Caltech), Josh Simon (Carnegie), Daniel Stern (JPL/Caltech), Louis Strigari (Stanford), James E Taylor (Waterloo), Joachim Wambsganss (Heidelberg), Randall Wayth (CfA, Harvard), Risa Wechsler (Stanford), Andrew Zentner (Pittsburgh)**

Contact: Leonidas A Moustakas, JPL/Caltech, 818-393-5095, leonidas@jpl.nasa.gov

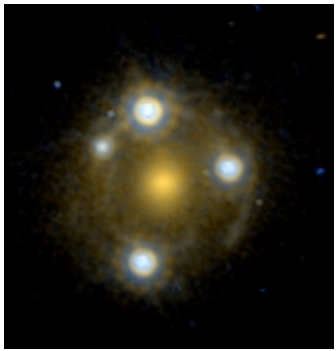

Shown is *HST* imaging of the galaxy-scale strong gravitational lens J0924+0219, where a distant quasar active galactic nucleus has been multiply imaged into four images. The flux, position, and light arrival time differences between all images encode a wealth of information about the relative distances and alignments involved, and about the gravitational potential in the lens itself. Small fluctuations in this potential due to small dark matter substructures result in *measurable* characteristic perturbations in the fluxes, positions, and arrival times. The framework for exploiting this is largely in place, and there is immense potential for significant inroads on the nature of dark matter this decade.

# Strong Gravitational Lensing Probes of the Particle Nature of Dark Matter

## Executive Summary

The puzzle of the nature of dark matter is one of the most profound questions of our era. The properties of a successful dark matter particle candidate are constrained by measurements of the dark matter cosmic abundance, Big Bang Nucleosynthesis, and four orders of magnitude in scale in the matter power spectrum, from the horizon size to scales of galaxies. There is a vast menagerie of perfectly acceptable options both within and beyond supersymmetric extensions of the Standard Model of particle physics. Each of these candidates may have scattering (and other) cross section properties that will satisfy all of the stated observables, but may lead to vastly different behavior at sub-galactic scales, and to significantly different "cutoff" scales, below which dark matter density fluctuations are smoothed out. The only way to quantitatively measure the power spectrum behavior at sub-galactic scales at distances beyond the local universe, and indeed over cosmic time, is through probes available in multiply imaged strong gravitational lenses. Gravitational potential perturbations by dark matter substructure encode information in the observed relative magnifications, positions, and time delays in a strong lens. Each of these is sensitive to a different moment of the substructure mass function and to different effective mass-ranges of the substructure. The time delay perturbations, in particular, are proving to be largely immune to the degeneracies and systematic uncertainties that have impacted exploitation of strong lenses for such studies.

**There is great potential for a coordinated theoretical and observational effort to enable a sophisticated exploitation of strong gravitational lenses as direct probes of dark matter properties. This opportunity motivates this white paper, and drives the need for: a) strong support of the theoretical work necessary to understand all astrophysical consequences for different dark matter candidates; and b) tailored observational campaigns, and even a fully dedicated mission, to obtain the requisite data.**

## Setting the Stage: Astrophysical Consequences of Dark Matter Particle Properties

In the current concordance Λ-dominated Cold Dark Matter ("ΛCDM") cosmology, we live in a universe dominated by dark energy, with nearly one quarter of the mass-energy budget in non-baryonic and cold, neutral, collisionless dark matter. At sufficiently early times in the universe, when the temperature is greater than the equivalent mass of the dark matter particle, dark matter is in equilibrium. For a given *self-annihilation* cross-section (which may depend on the particle mass), as the universe expands, the annihilation rate at some point drops below the Hubble expansion rate. At this point, annihilations become rare, and the abundance of dark matter is frozen to the densities currently observed. With a larger *scattering* cross section, dark matter remains in an approximate thermal equilibrium until the expansion forces these interactions to become rare as well. The temperature at this thermal-decoupling event

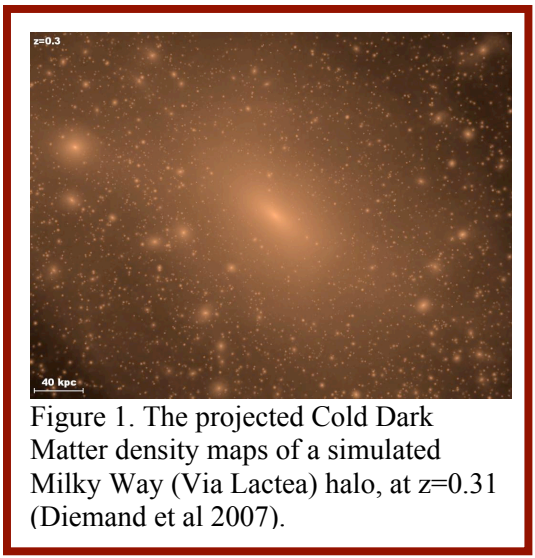

Figure 1. The projected Cold Dark Matter density maps of a simulated Milky Way (Via Lactea) halo, at z=0.31 (Diemand et al 2007).

# Strong Gravitational Lensing Probes of the Particle Nature of Dark Matter

sets both the particle phase space density of dark matter, and the free-streaming scales of the smallest structures. This in turn directly causes the truncation of the power on small scales, down to the free-streaming-based fundamental cutoff. If dark matter is a "classic" supersymmetric weakly interacting massive particle (WIMP) with a self-annihilation cross section comparable to the electroweak cross section, then at small scales, the matter power spectrum is predicted to be (largely) self-similar down to an effective cutoff around $10^{-6}$ $M_{sun}$ (Schmid et al 1999; Green et al 2005; Loeb & Zaldarriaga 2005).

This model is remarkably successful at matching observations from scales that span the horizon length all the way down to scales of ~1 Mpc, or perhaps even ~100 kpc. The smallest of these scales have been probed through power spectra of Lyman-alpha forest absorbers at redshifts above $z>2.4$, and have already provided some powerful constraints on some properties of alternative dark matter candidates (Seljak et al 2006). At yet smaller scales, (equivalent to the tiny substructure within galaxy-scale dark matter halos; Figure 1) there have been persistent observational challenges to the WIMP-based cold dark matter expectations. This tension is predominantly found in the numbers, phase space densities, and density profiles of dark-matter-dominated halos in the local universe (including within our Galaxy), and has motivated enormously sophisticated new observational campaigns of the objects of interest, and allowed for a much broader view of what may be the constituents of dark matter. This interest has been compounded by recent high-energy observations (Adriani et al 2008) that may be related to dark matter annihilation processes (Arkani-Hamed et al 2009), and the debate (and the data) will undoubtedly be sharpened even more this decade.

The key here is that the most straightforward expectations from supersymmetry are not sacrosanct. There is great richness in plausible non-baryonic particle origins and properties that, while preserving all the larger-scale successes of our current model, have very different thermal decoupling times (SuperWIMPs [Feng et al 2004], sterile right-handed neutrinos [Dodelson & Widrow 1994]), or entirely different (even non-thermal) production mechanisms (Axions, Universal Extra Dimensions, or gravitinos from Supergravity). Each of these may produce perfectly acceptable matches to the horizon-to-galaxy scale observations, but with a broad range of departures at the smaller scales, resulting in different particle phase space densities and dark matter halo density structures, as well as different matter power spectrum shapes and cutoff scales. An example is graphically shown in Figure 2, where the small-scale consequences of a 3 keV dark matter particle are immediately clear. There are two further very important points to make. First, dark matter could be a mixture of more than one type of particle, forcing any direct or indirect detection to paint an incomplete picture. Second, many of the plausible dark matter candidates have particle-interaction properties that may make them simply unobservable in any direct or indirect detection experiments. Both of these cases make an even more forceful case for direct measurements of the small-scale matter power spectrum, *over cosmic time*, as an indirect but key probe to the nature of dark matter. We present a powerful probe that, with sufficient resources and dedication, will in the coming decade enable quantitative constraints on the power spectrum at these critical scales.

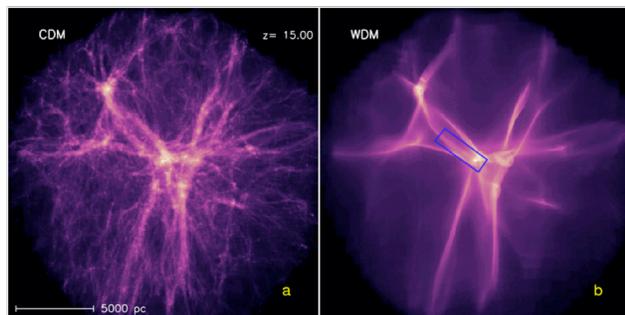

Figure 2. A comparison of small-scale structure from standard Cold Dark Matter (left) vs Warm Dark Matter with a 3 keV dark matter particle (Gao & Theuns 2007).

## Strong Lensing Probes of the Matter Power Spectrum

Strong gravitational lensing is a phenomenon by which a sufficiently high mass concentration will focus the light from a more-distant aligned object into multiple images of the same source. For this discussion, we are concentrating on multiply imaged time-variable compact sources, such as active galactic nuclei which are lensed by galaxy-scale lenses.

All the salient features of a strong lens may be described by the two-dimensional surface of the arrival times of photons from the distant, lensed, source. This time delay surface is derived from the relative positions and distances of the observer, the lens, and the source, and the shape of the smooth lensing potential. The arrival times depend directly on the lensing potential. By Fermat's principle, the positions where images will form are given by the extrema of the time delay surface, and so depend on the first derivative of the potential. The magnifications, or fluxes, of the images depend on the second derivative of the potential. Since the true intrinsic properties of the source's position and flux are not known, generally one measures the ratios of these quantities between the images in a given lens.

Dark matter substructure in a lensing dark matter halo may conveniently be described by a differential mass function *dN/dm*, which may be derived by the fluctuation power spectrum on the same scales. The dark matter substructure in the halo leads to small but important potential fluctuations that can be manifest in perturbations in each of these quantities, which should be measurable with the appropriate observations in the right samples of lenses. Naturally, there are both practical and theoretical limitations that must be overcome or accounted for in each case. The key to both the practical and the theoretical aspects of this problem lie in the derivative-order dependence of each quantity (Keeton & Moustakas 2009 [KM09]).

We now expand on this point, beginning with the case of **relative flux perturbations**. As magnifications are dependent on the second derivative of the potential, even very small dark matter substructure fluctuations may result in order-unity flux differences. For the same reason, these perturbations correspond to a measure of the $\int(m)dN/dm$ moment of the dark matter mass function. This means that these perturbations are dominated by a small number of the nearest dark matter substructures to each image, nearly independently of the mass of the substructure. Thus, relative flux perturbations can be described as quasi-local tracers of the total mass in dark matter substructure (Table 1). In practical terms, stellar microlensing by stars in the lensing galaxy can also effect order-unity fluctuations, which vary on the timescales of weeks to years. The stochastic contribution from microlensing can be mitigated either through long-term monitoring of the lens, or through flux-measurements that are not affected by stellar-sized micro lenses. The microlensing must be accounted for or avoided entirely, e.g. through high-resolution imaging in the radio (Dalal & Kochanek 2002; Dobler & Keeton 2006) or the mid-infrared (e.g. Chiba et al 2005), or with spatially-resolved spectroscopy (Moustakas & Metcalf 2003; Metcalf et al 2004). The potential of this probe has been limited so far by the small number of applicable

**Strong Gravitational Lensing Probes of the Particle Nature of Dark Matter**

targets (e.g. only 6 radio-bright four-image lenses). Even so, useful dark matter particle limits have been placed (e.g. Miranda & Maccio 2007). With *JWST*, mid-infrared flux perturbation measurements which by-pass the microlensing "noise" issue will be possible for one or more orders of magnitude more lenses. Finally, the connection between potential perturbations and relative flux perturbations requires extensive simulations for each targeted lens. The exercise is challenging, but tractable, especially as there are recent developments on the mathematical theory front (e.g. Petters et al 2008). The statistical representation from large number of lenses will lead to ever-better fidelity in the measurement of dark matter particle properties.

In a relatively smooth lensing potential, the relative positions of the lensed images are fairly set, and substructure can cause **astrometric perturbations** (e.g. Metcalf & Madau 2001). These typically manifest at sub-milli-arcsecond levels, and so require extremely high angular resolution, typically achieved by VLBI observations of radio-bright lenses. This has so far limited the number of lenses that could be targeted, though the theoretical promise continues to be explored (Chen et al 2007, Vegetti & Koopmans 2009). Because these perturbations are sensitive to the substructure mass function in a somewhat top-heavy fashion (KM09 & Table 1), they probe an intermediate substructure mass regime, and are sensitive to features at a larger impact radius around each lensed image.

We continue with the consideration of **time delay perturbations**. These depend directly on the potential, and as shown in KM09, dark matter potential fluctuations lead to gaussian perturbations in the time delays in the sense of $\sigma_{dm} \propto (f_{dm} m)^{1/2}$, i.e. with a dependence on the fraction of mass in substructure and on a characteristic mass of the mass function, $m = \langle m^2 \rangle / \langle m \rangle$. These perturbations measure the $(m^2)$ moment of the dark matter mass function, $\int (m^2) dN/dm$, and hence are more sensitive to the high-mass end (Figure 3 & Table 1). The direct dependence on the potential (and its fluctuations) leads to the time delay perturbations being due to the sum of small effects from many (distant) clumps, making this a much more global tracer of dark matter than relative fluxes. Spectacularly, if the time delay *ratios* within a given strong lens are considered, the usual host of degeneracies associated with strong lenses (including the mass-sheet degeneracy) are largely removed, leaving geometry (from the relative positions and distances) and the potential perturbations as the dominant lensing observables (KM09). For typical CDM levels of substructure, the time delay fluctuations are expected to be on the order of ~1%, though with a large spread. Presently, time delays are measured to precisions approaching several percent (Oguri 2007), though Congdon et al (2009) have demonstrated that even now there are several strong lenses with anomalous time delays, which require the presence of potential fluctuations. (A detailed comparison against CDM substructure predictions is work in progress.)

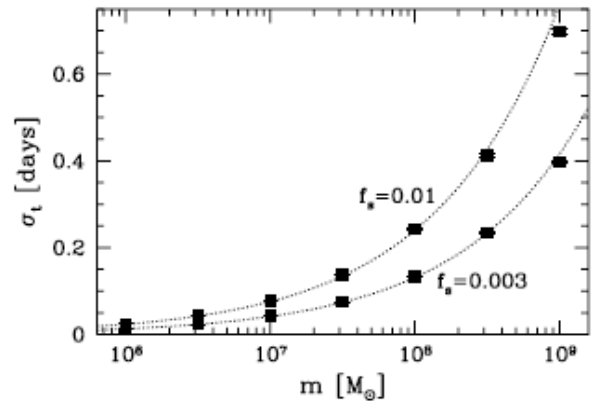

Figure 3. From KM09. The time delay scatter as a function of subhalo mass. The points show results from Monte Carlo simulations, with statistical errorbars from bootstrap resampling. The dotted curves show the analytic scaling for $\sigma_{dm}$.

# Strong Gravitational Lensing Probes of the Particle Nature of Dark Matter

| Strong lensing dark matter substructure probe | Dark matter mass function moment dependence | Dark matter substructure mass range sensitivity | Sensitivity to area around each lensed image | Sensitivity to the internal structure of substructure | Main observational challenges |
|---|---|---|---|---|---|
| Time delays | $\left(\langle m^2 \rangle / \langle m \rangle\right)^2$ | High mass (<$10^9$ M$_{sun}$) | Long-range | Little | High time domain precision |
| Relative positions | $\left(\langle m^2 \rangle / \langle m \rangle\right)^{3/2}$ | Intermediate to high mass | Intermediate | Modest | High astrometric precision; lens modeling |
| Relative fluxes | $\left(\langle m^2 \rangle / \langle m \rangle\right)$ | Full mass range | Quasi-local | Sensitive | Microlensing; lens modeling |

**Table 1** – Some of the salient characteristics of the three strong lensing based probes of dark matter substructure are presented.

In even a single strong lens, then, it is possible to make high-precision measurements of different moments of the dark matter mass function, which are weighed to different mass-ranges. What is required to enable all of these *simultaneously* would be high angular resolution imaging at long wavelengths (for microlensing-free relative fluxes and reasonably precise relative positions), combined with densely populated high signal to noise light curves of all images in the lens, towards high precision relative time delays. Because of the relatively small number statistics involved in these perturbations, it is highly desired to compile such measurements on a large sample of strong lenses, especially for exploring the dark matter substructure properties over a range of redshifts, and cosmic time.

***The synthesis of all three strong lensing probes in large samples of lenses will produce a direct measurement of the dark matter mass function across decades of mass-scales and over time.***

## Required Observations

The requisite measurements may be achieved by combining observations from a variety of platforms. First and foremost, large useable samples of lenses must be found – presently, fewer than one hundred useful candidates are known. The advent of large time-domain surveys in the optical and in the radio are expected to yield thousands of lensed and variable active galactic nuclei. These possible developments are explored in more detail in the White Papers contributed by L. Koopmans and P. Marshall. Space-based wide-field imaging observatories, such as *JDEM* or *Euclid*, would also produce hundreds to thousands of new useful lenses. These large new samples would be the target of dedicated follow-up in the context of the science advocated herein.

The separation between lensed images is typically between one-tenth and one arc-second, and high-fidelity relative flux measurements require a photometric precision of a few percent. This may be an achievable project for next-generation Adaptive Optics systems in large telescopes, working in the near-infrared. At present, many, if not most, known targets are beyond the practical reach of even the Keck-based AO platform. In the coming decade, *JWST* mid-infrared

**Strong Gravitational Lensing Probes of the Particle Nature of Dark Matter**

imaging will comfortably achieve the long-wavelength high-resolution and high signal to noise photometry required.

The outstanding probe left is in the time delay perturbations. As demonstrated in KM09, the real power of time delay ratios can be fully realized only with a large ensemble of lenses with well-measured time delays. Time delays in galaxy-scale lenses typically range from hours to weeks. At present, time delay uncertainties are tied to the limitations inherent to ground-based observations, and the best are known ~1 day (Oguri 2007). An order of magnitude improvement is desired. As an example, if uncertainties are improved to ~0.5%, they would yield constraints on substructure at the level of 0.3–0.5 dex in a single given four-image lens. The better the time-delay precision, the better the fidelity of the substructure measurements. Achieving time delay precisions for large samples of galaxy-scale strong lenses is beyond the capabilities of ground-based platforms. A one-meter class space-based observatory with a multi-band imager, dedicated to the monitoring of 100 or more variable strong lenses, would comfortably achieve this objective. A mission concept for this is being developed: the Observatory for Multi-Epoch Gravitational Lens Astrophysics (OMEGA; Moustakas et al 2008).

**Bridging the 2010s Observations to Conclusions about the Nature of Dark Matter**

If a sample of ~100 four-image lenses can have sub-percent-level relative fluxes, sub-milli-arcsecond precision image positions, and relative time delays measured to better than ~0.5%, that should yield quantitative substructure mass-function measurements at levels of better than 20% across a large range of substructure masses. This will be an important result, but casting it in the context of explicit predictions from different dark matter candidates will realistically require a Bayesian prior approach. For ranges of dark matter candidate properties, robust theoretical predictions should be made for the behavior of the small-scale power spectrum, which can then be a posteriori compared against the observational constraints. This will be a great challenge in the first few years of this decade. Transfer functions towards quantitative matter power spectrum predictions can be made by solving the relevant Boltzmann equation and following the evolution of structure with time. However, the observed regime is by construction non-linear and within massive collapsed objects, where the environment is potentially volatile to small-scale structure, through tidal forces from both dark matter and baryonic components in each galaxy,

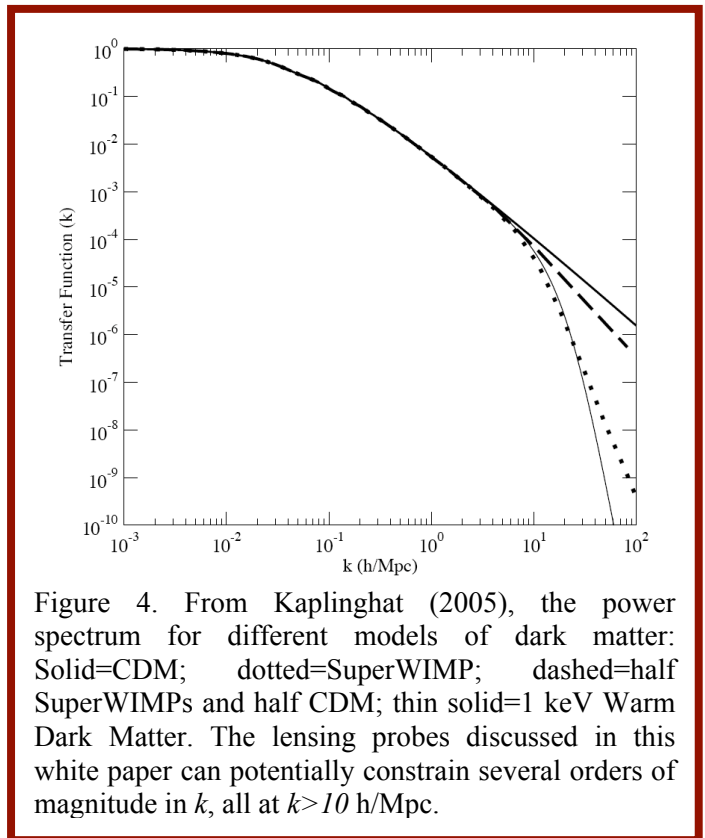

Figure 4. From Kaplinghat (2005), the power spectrum for different models of dark matter: Solid=CDM; dotted=SuperWIMP; dashed=half SuperWIMPs and half CDM; thin solid=1 keV Warm Dark Matter. The lensing probes discussed in this white paper can potentially constrain several orders of magnitude in $k$, all at $k>10$ h/Mpc.

**Strong Gravitational Lensing Probes of the Particle Nature of Dark Matter**

and more (e.g. Metcalf & Madau 2001; Zentner & Bullock 2003). This will require detailed N-body simulations and semi-analytic modeling to be systematically tackled, far beyond what has been done to date, for a very broad range of *non-classic dark matter candidates*.

### Recommendation

**We discuss how a synthesis of complementary measurements in time-variable strong gravitational lenses may function as powerful probes on the behavior of dark matter substructure as a function of redshift, at scales that can discriminate between competing dark matter particle models. First and foremost, these observations will be able to** *directly test the CDM ansatz***, and** *will lead to inferences on several key physical properties of the constituent or constituents of dark matter***. These measurements are** *independent from* **and** *complementary to* **any direct and indirect detections of dark matter that may materialize in this decade.** In fact, if the fundamental properties of dark matter prevent it from producing verifiable detectable signals, the proposed observations will be a powerful and largely *unique* probe of the nature of dark matter through its smallest-scale structure.

In the coming decade, there promises to be an explosion in the discovery of strong gravitational lenses through a variety of means, predominantly through ground-based time-domain searches. The full exploitation of these lenses for the purposes of dark matter studies will require extremely high-quality follow-up observations, for two primary reasons. First, the separation between images in a multiply-imaged active galactic nucleus is much smaller than an arc-second. Second, the time delays between these small-separation images need to be measured to very high precision. An observational campaign that achieves these measurements, whether in a single observatory or by coordination across platforms will lead to robust measurements of the matter power spectrum over some four decades of sub-galactic scales, as a function of redshift. The most important, and yet least anticipated measurements are in the high-angular resolution high precision time delay measurements of large samples of such lenses. These goals can be accomplished through the OMEGA mission concept, a 1-meter class space-based observatory that would be dedicated to the constant monitoring of a large sample of lenses.

### REFERENCES & FURTHER READING

**Further reading available at http://www.its.caltech.edu/~leonidas/OMEGA**